# Adding Methodological Testing to Naur's Anti-formalism

Steven Meyer
Tachyon Design Automation Corp., San Francisco, CA, USA
**smeyer@tdl.com**

## 1. Introduction

In late 2011 Turing Award winner Peter Naur published an extremely interesting book (Naur[2011]) in the form of an interview. The book strongly criticizes formalist computer science. Naur's characterization of computational thinking is superior to previous criticisms because he understands that computer science has become "cognitive information processing". Naur states "mental life during the twentieth century has become entirely misguided into an ideological position such that only discussions that adopt the computer inspired form" are accepted. (Naur[2007], 87). This formalist conception of the nature of human cognition now dominates not just theoretical computing, but also computer program development, social sciences, philosophy of science and is even currently advocated as a replacement for physics (Aaronson[2005], p. 1).

The book *Conversations - pluralism in software engineering* is particularly valuable because Naur provides background knowledge and explains the evolution of his current thinking. In addition, the interviews make Naur's publications more easily accessible by giving URL's for some of his more recent works and by providing detailed references for his earlier works during his amazing 60 year career as a leading founder of computing.

My goal in this paper is to supplement Naur's criticism by describing a different philosophical tradition that adds methodological testing to Naur's critical philosophy. Naur's criticism of AI and the mechanical cognitive metaphor is unrivaled, but in my view Naur does not provide any prescriptive guidance to scientists and philosophers for choosing between alternative methodologies and does not provide assistance so that scientists can test research programmes.

After summarizing Naur's criticism, I discuss a related philosophical tradition that when added to Naur's, allows methodological testing of computer dataology and scientific methods and theories. The tradition is called "the methodology of scientific research programmes" (MSRP) (Lakatos[1970]) and was developed by philosopher Imre Lakatos as an antidote to computer inspired forms of thought. MSRP is the culmination of anti-formalism through the study of method started by the founders of modern physics and continued by the Vienna Circle. Next, four different criticisms of structured programming are discussed including my 1970s criticism that explicitly attempted to apply MSRP. The remainder of the paper discusses anomalies and problems that need to be studied with this new methodology in a post computer inspired methods world.

## 2. Naur's Criticism of Mechanical Cognitive Information Processing

Naur recognizes the importance of programmer specific problem solving. In Naur[2011] page 30, the interviewer asks "... you basically say that there are no foundations, there is no such thing as computer science, and we must not formalize for the sake of formalization alone." Naur answers, "I am not sure I see it this way. I see these techniques as tools which are applicable in some cases, but which definitely are not basic in any sense." In criticizing structured programming, Naur states: (p. 44) "The programmer has to realize what these alternatives are and then choose the one that suits his understanding bests. This has nothing to do with formal proofs."

Naur understands the importance of terminology and criticizes computer science's attempt to justify a theory by giving it a persuasive name. Naur believes that calling computing "computer science" is misguided and suffers from the fallacy of proof by naming (Naur[2005], 208). Naur invented the better term dataology that I use in this paper (Naur[2007], 86).

Naur states (Naur[2011], 67) that "program development is about building up a certain understanding, a theory, its not about creating program text." Naur explains that there are no formal principles for the building and uses this building a theory metaphor in his dataology methodology also. Naur relates that program development can be viewed as programmers injecting their knowledge into programs (p. 50). The theory building metaphor is accompanied by Naur's superior cases studies and constructed models in dataology and in science (Naur[1995], 270).

## 3. Discussion of Naur's Philosophy

Naur advocates that science and philosophy should be based on descriptions. In his paper "Computing as science", Naur lists eight aspects of various fields of learning (Naur[2005], 208-210). For example, the eighth circumstance is "the influence from computers on how human mental activity is described. During the last 40 years many psychologists seem to talk about human beings as information processors as a matter of course" (p. 210). Naur's solution is a philosophy of descriptions that are internally coherent. (p. 212).

In my view, some minor problems with Naur's philosophy are that it is to narrow and too much based on what Wittgenstein called pointing and does not deal with theoretical content. In addition, I think Naur's historical analysis is not quite correct. Problems arise when Naur's philosophy is applied to scientific practice because methods and theories are not testable. There is no rational method for deciding between alternatives: between alternative data representations, different program verification methods and different theories or research programmes.

I do not believe as Naur claims that Karl Popper's logic of scientific discovery was an important part of scientific practice. Naur's statement that "since the middle of the 20th Century, the discussion of what science is has been dominated by Popper's notion" (Naur[2005], 211) is wrong.

The actual history is that around 1900, the founders of modern physics realized that science needed objective ways of testing competing theories because the stable Newtonian universe was breaking down. Particularly Max Planck and Albert Einstein realized that objective ways to compare competing theories that often had almost nothing in common were needed. The beginning of quantum physics connects to dataology because Planck's black body radiation calculation depended on the fact that there is only a countable number of the separate non-interacting oscillators (Bohm[1951], 15). Their students and colleagues then in the 1920s and 1930s formed the Vienna Circle which criticized metaphysics and attempted to provide rational ways of testing competing theories. Karl Popper was not part of the Vienna Circle. After WWII Popper was ideologically popular, though. Popper's conception of falsifiability was not accepted by the Vienna Circle in the 1930s and was not popular among scientists in the 1950s before Lakatos. Scientists viewed falsification as too simple. Although, sometimes simple falsification provides assistance to scientists. In the 1930s, Vienna Circle founder Otto Neurath criticized Popper's falsificationism (Neurath[1936]) in a letter dated 4.2.1936 to Karl Popper by stating "Einstein and Planck believed in schools." (modern term for "school" is "research programme"). Also, Naur does not use the term "logic" in the sense it is used by scientists and philosophers. The philosophical as opposed to mathematical meaning of the term "logic" follows the traditional intuitive sense of "consistent internal relationships" that goes back to Greek philosophers.

During the late 19th and 20th century, mathematics split into pure and applied parts

(Naur[2005], 209). This splitting was opposed by both classical mathematicians such as Hilbert and physicists but occurred anyway. The main change was elimination of any connection of mathematical entities from physical reality and replacing mathematics with the study of abstract structural properties. See Chandler[1982] for a detailed history of this change.

Also, I think Naur's characterization of science as: "Every science is a matter of description (Naur[2011], 85) actually is the same philosophical theory as the later Ludwig von Wittgenstein's philosophy of mathematics (Wittgenstein[1939]). The idea of mathematics as "describing" was Wittgenstein's response to the the split into pure and applied mathematics and the acceptance of structure abstracting formal set theory. I think Wittgenstein would have called Naur's describing as pointing. Paul Feyerabend gave his impression of Wittgenstein in 1952 by writing: Wittgenstein emphasized "the need for concrete research and his objections [were] to abstract reasoning ('Look don't think!')" (Feyerabend[1978], 115).

I think philosopher of mathematics Juliet Floyd would characterize Naur's philosophy as similar to Feyerabend's, Kuhn's and Popper's. She writes:

*Feyerabend, like Popper, missed the boat. They each missed the multifariousness of the ways in which modern formal logic would serve as a new lens for philosophy, illuminating and distorting its questions in new kind of ways (Floyd[2011], 111).*

I would use the term "quasi-empirical scientific testing of methods" for "philosophy" above, but otherwise agree. See my 2011 IACAP paper that discusses logical truth from Lakatos' quasi-empirical research programme perspective (Meyer[2011]).

## 4. Lakatosian Methodology of Research Programmes (MSRP)

Feyerabend describes MSRP in this way:

*He [Lakatos] admits that existing methodologies clash with scientific practice, but he believes that there are standards which are liberal enough to permit science and yet substantial enough to let reason survive. The standards apply to research programmes, not to individual theories; they judge the evolution of a programme over a period of time, not its shape at a particular time; and they judge this evolution in comparison with the evolution of rivals, not by itself. (Lakatos[1999], 116)*

The missing part of Naur's characterization of computing and science is that it is irrational because it allows acceptance of whatever the current dominant academic clique believes and facilitates elimination of all competing theories. This is exactly the situation that occurred during the 1970s as computer science became mechanical cognitive information processing (AI). The dominance of AI and object oriented formal program verification happened as Naur has documented without any debate let alone any scientific testing of competing research programmes. Basically, current computing became what is it because of mob rule by influential and well funded academics (Naur[2007], 87). The remainder of this paper adds research programme testing to Naur's characterization of dataology and science.

## 5. Methodological Advantages of my Falsification of Structured Programming

The importance of adding methodology testing to Naur's philosophy can be seen by considering the history of the disproof of formal program verification. I used the MSRP theory to falsify structured programing but my MSRP based disproof was ignored so that it took at least 30 more years for skeptics to prevail.

There are four different research programmes criticizing structured programming. By structured programming here I mean not just Naur's "idea's of so called structured programming" and "claims for so called formal specifications of programs" (Naur[2007], 86) but also belief in the supremacy of mathematical logic and set theory. This belief in absolute formalism convinced adherents that it was only a matter of time before artificial intelligence surpassed human intelligence and before all mathematics was mechanized. Structured programming was the first human involved step of this inevitable formalization of everything.

The four criticisms are:

1. My MSRP based criticism falsified structured programming by showing that a defining example by Edgser Dijkstra failed. The algorithm is inefficient in spite of Dijkstra claim. Dijkstra saved correctness by finding an alternative English language parsing of the text describing the refinement (Meyer[1983], 4-9 and 11). My criticism used methodological testing from MSRP to falsify the hard core of structured programming by disproving a defining instance of the research programme because examples claimed to define a theory must be its hard core.

2. Naur's "programming as a human activity" with the hard core of programming as development of "a certain kind of [human] understanding" (p. 86).

3. Demillo[1979]'s program verification must fail because mathematical proofs are accepted from social processes, and no comparable social processes can take place among program verifiers (p. 271). As Feyerabend would put it "no mob can be found to rule" (Lakatos[1999], 117). I view the DeMillo[1979] argument as a Naur style case of "ideological suppression of scientific discussions" [Naur[2007], 87) because the DeMillo[1979] authors were referees of my structured programming paper (situation is documented in Meyer[1983] especially p. 23) but did not reference my paper. The ideology comes from Demillo[1979]'s defending mathematical formalism against my MSRP research programme testing criticism by jettisoning program proving in order to save formalism.

4. Fetzer's criticism claims that algorithms can be proven because they are axiomaticized formal structures but program are "casual models" that can not be proven (Fetzer[1988], 1048). Fetzer's criticism requires accepting certain properties of formal structures.

I claim my criticism is the best of the four because it is objective and defends rationality using MSRP for its falsification. It scientifically and objectively falsified structured programming. Naur's criticism depends on human psychology (I think he would claim this is a positive) so it can not be tested and there is no progress since the "laboratory behavior" of one person compared to another is hard to test. DeMillo[1979]'s criticism is irrational because it basically calls for mob rule (social processes). Fetzer[1988]'s criticism depends on questionable beliefs about formal systems. For example, using Fetzer's argument, proofs in temporal logic are invalid but proofs in ZF set theory are valid because in temporal logic events in times are followed by events at a later time (this is usually what is meant by "caused").

## 6. Toward Dataology Academic Study of Computing

Naur's recent conversations book along with his 2007 Turing award lecture have made huge strides in allowing study of computing in a scientific way. I think it now makes sense to think about how academic dataology departments should be organized and what they should study.

### 6.1 Ideological Suppression Issue Must be Solved

Before dataology academic departments can be established, scientific independence must be established. As Naur puts it: "Other issues of science and scholarship imposed themselves upon my attention in the form of ideological suppression of scientific discussions of computing and human thinking." (Naur[2007], 87-88). Actually, the issues that Naur discusses have happened in the US previously. See Thorstein Veblen's late 19th century expose of the economic reasons for ideological suppression in US universities in his book *The Higher Learning in America,* Veblen writes "The need of university prestige ... pushes the members of the staff into a routine of polite dissipation, ceremonial display, exhibitions of quasi-scholarly proficiency and propagandist intrigue." (Veblen[1918], 124)

I can add additional suppression experiences that are more organizational than Naur's. I attended the UC Berkeley computer science department after receiving a B.S. degree from Stanford in physical science (individually designed major). I attended the Berkeley CS department that was an independent department that connected with physics because computing began as numerical analysis. I looked for a computer science department because of advice from William Shockley to avoid going to graduate school in an engineering department from his experiences. I knew Shockley from attending his Freshman seminar. I shared an office with Diane McIntyre who had also studied at Stanford. Both of us were educated in an environment that encouraged study of concrete problems because of George Polya. Naur quotes Polya: "start with concrete examples and then get the knack of it." (Naur[2011], p. 92). In my case, I was even encourages at Stanford to avoid learning logic and set theory.

Both of us were thriving from the Polya influence and were allowed to study what we wanted until the CS department was taken over by the EE department and all assistant professors were fired. I had passed my orals with adviser Jay Earley and also James H. Morris. I ran into ideological suppression when I became convinced that automatic programming couldn't work, and since I had been attending Feyerabend's seminar decided to try to somehow show that my belief was right. That is the reason for my MSRP falsification of Dijkstra's Dutch National Flag problem analysis (really sorting with only 3 values) (Dijkstra[1976]). The ideological suppression followed a letter from Dijkstra that caused the then EE tenured faculty to prevent refereeing of my paper. The long story is documented in Meyer[1983], 4-21. I ended up being right, but it took almost 40 years.

Another ideological suppression issue involves sociology. Sociologists seem to be the strongest supports of the humans as information processors research programme. For example, the 2001 book *Mechanizing Proof* by Donald MacKenzie (MacKenzie[2001]) totally mis-represents Naur's views. Although the book was published in 2001, Naur is given only two paragraphs on his programming methods. Mackensie writes: "He was a central member of the influential group of "mathematicizers" within computer science, but the mathematics he sought to apply was ordinary informal mathematics." (p. 49) This seems a rather strong understatement of Naur's criticizing to me. McKensie does write: "Naur, indeed, was later to become an explicit critic of excessive formalism. (p. 49). Where is the individual choice mention? Also, McKensie's book barely discusses the Unix operating system that is very widely used in areas involving risk and trust (p. 98) and does not mention the most widely used Linux system and Linus Torvalds who developed a personal program development method.

### 6.2 Two Organizational Proposals

First, there needs to be independent datalogy departments that will be able to allow multiple competing research programmes to coexist. This has not been possible after computer science departments were moved into engineering schools and EE departments. Engineering is aimed building and improving repeatable processes.

Second, there needs to be a journal like the old Communications of the Association of Computing Machinery (CACM) that quickly publishes shorter results and algorithms with minimal review for relevance. Currently, CACM is a popular journal not intended for scientists. The change to many specialized journals starting in the mid 1970s was a mistake because it strengthened an ideological based excluding review process. Another possibility would be to have a journal similar to the Physics Review Letters.

### 6.3 Study Areas for Human Centric Dataology

There are a number of areas that once normal scientific study is restored could be studied in dataology departments.

1. **Reconnect with 19th century efforts to access infinity.**
   Mathematics seems unable to study issues involving infinity that do not involve current standardized axioms and set theory. Under Naur's method, the difference between potential and actual infinity could be studied as competing research programmes. An area of study similar to Naur's universal Turing machine analysis might be studying the difference between the 1859 Dedekind cut definition of real numbers (Goldrei[1998] 8-17]) and the later Cantor definition that requires an equivalence class algorithm (pp. 19-21).

2. **Research program competition among different groups of axioms.**
   Swiss mathematician Paul Finsler, who studied at Goettingen during the Hilbert era, argued that axiom selection was not just conventionalism, but should be tested. Compare Finsler's argument that the independence of the continuum hypothesis is an objective true/false question similar to the independence of the parallel line axiom in geometry (Finsler[1969]) with Dana Scott's proof of the independence of the continuum hypothesis (Scott[1967]). See also Finsler[1996]. Along the same lines, it seems to me that Risch's closed form indefinite integration algorithm (Risch[1969]) is somehow wrong because it omits too much human mental life. Smolin's discussion of problems with the mathematics of physical fields in high energy physics illustrates the limitations. (Smolin[2006]).

3. **Compare Penrose physicist's description of memory with Naur's synapse model.**
   The way physicists think is different from dataologist thinking. Roger Penrose's disproof of AI by positing quantum microtubule memory (Penrose[1994], 366-367) should be compared with Naur's Synapse State model (research programme) (Naur[2011] 94-116).

4. **Concrete complexity algorithms discovery from efficiency proof analysis**
   Currently, algorithms in the concrete computational complexity area are discovered by analyzing the combinatorial efficiency proof using standard assumptions and counting conventions. This seems to me to need testing since it leaves out Naur's human individuality problem solving. Possibly new research programmes with different rules for counting steps need to be discovered so there can be competition with current methods. Lakatos, in his book *Proofs and Refutations* identified problems with discovery through proof analysis (Lakatos[1976], 52-56 and 142).